\documentclass[11pt]{article}
\usepackage[a4paper,margin=1in]{geometry}
\usepackage{amsmath,amssymb,amsthm,mathtools}
\usepackage{hyperref}
\usepackage{graphicx}
\usepackage{booktabs}
\newtheorem{theorem}{Theorem}[section]
\newtheorem{proposition}[theorem]{Proposition}
\theoremstyle{remark}\newtheorem{remark}[theorem]{Remark}
\title{Computing Smooth Surfaces via Nets of Geodesics}
\author{Tom Gilat\\
    Faculty of Mathematics\\
   Technion -- Israel Institute of Technology\\
    \texttt{tom.gilat@technion.ac.il}
}
\date{November 2026}
\begin{document}
\maketitle
\begin{abstract}
We study the reconstruction of smooth surfaces from networks of spatial curves prescribed as geodesics. Motivated by applications in computer-aided geometric design and the perception of three-dimensional shape, we propose a variational approach minimizing the integral of squared Gaussian curvature over surfaces spanning geodesic contours.

Using isothermal coordinates, we derive the Euler--Lagrange equation for the conformal factor of the metric and the associated Dirichlet and Neumann boundary conditions. We also consider a biharmonic approximation obtained by linearizing the nonlinear equation.

Numerical experiments are performed on four quadrilateral cells of a geodesic network constructed on an ellipsoid. We compute approximate isothermal coordinates, solve the biharmonic boundary-value problem, and compare the resulting curvature with that of the reference surface. These experiments consist of an important step towards a complete framework for the computation of smooth surfaces from geodesic curves.
\end{abstract}

\section{Introduction and motivation}
The problem considered in this work is the construction of a smooth surface in three-dimensional space from a prescribed network of curves, with the requirement that these curves be geodesics of the resulting surface. This problem arises naturally in computer-aided design (CAD) and geometric modeling, where the construction of smooth surfaces from geometric constraints is of considerable practical importance. The smoothness of the resulting surface is significant both aesthetically and in applications such as automotive design, architecture, and industrial manufacturing.

An additional motivation comes from the observation that artists and illustrators frequently use networks of curves to convey the geometry of three-dimensional objects. We hypothesize that some of these curves may be understood as projections of geodesics on the depicted surfaces. An intriguing example is Rembrandt's drawing of an elephant, in which several networks of curves appear to describe the animal's three-dimensional form. These curves may be interpreted as representations of virtual geodesics on the elephant's skin.

\begin{figure}[htbp]
\centering
\includegraphics[width=0.68\linewidth]{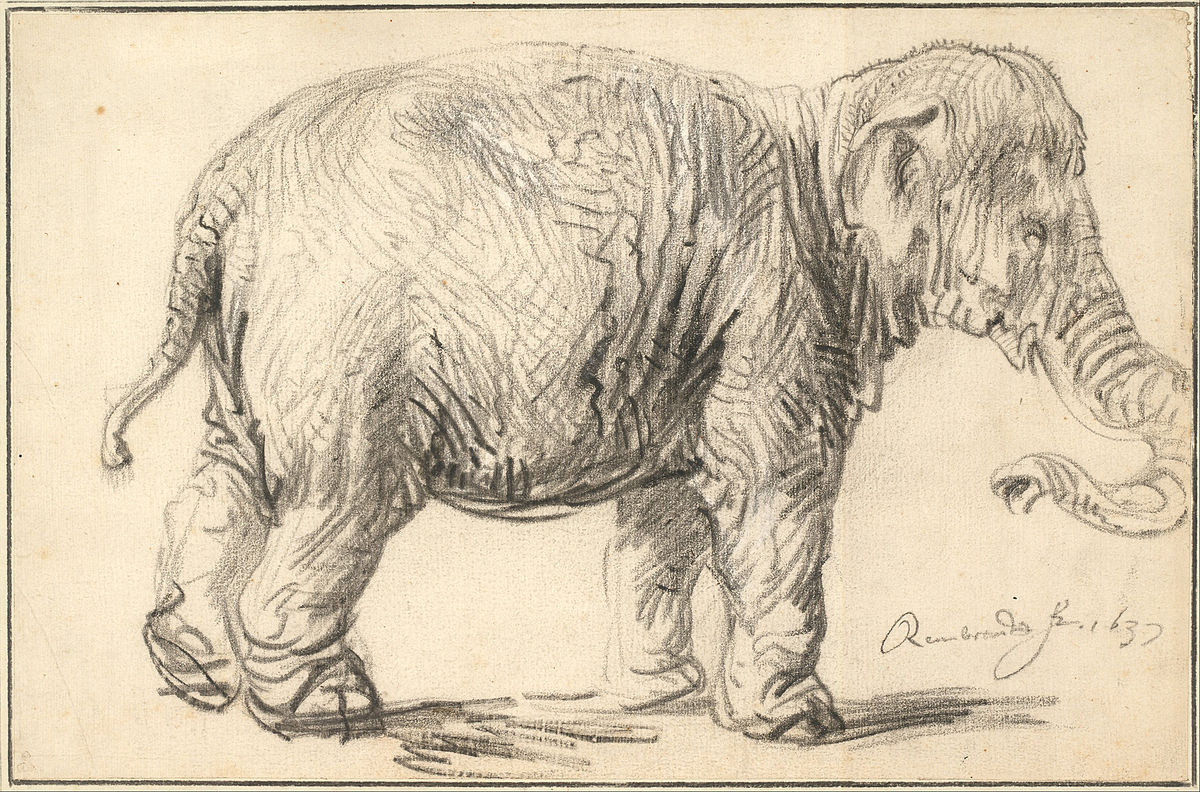}
\caption{An Elephant, Rembrandt, 1637 (source: Wikimedia). Networks of curves appear in several regions of the drawing, including the head and body. Their possible interpretation as projections of surface geodesics is part of the motivation for this work.}
\label{fig:elephant_rembrandt}
\end{figure}

This interpretation is related to research in visual perception. In particular, David Knill investigated how projected surface contours contribute to the perception of three-dimensional shape \cite{Knill92}. This connection suggests that geodesic networks may provide a natural geometric representation for both the perception and the reconstruction of surfaces.

From the mathematical viewpoint, the problem raises the question of how to select a surface among those compatible with prescribed geodesic boundary curves. We consider the minimization of the total squared Gaussian curvature,
\[
\mathcal E(S)=\int_S K^2\,dA,
\]
as a variational principle for selecting such surfaces. The resulting problem combines differential geometry, nonlinear partial differential equations, and numerical surface reconstruction.

Our approach begins by studying a single surface patch bounded by a piecewise smooth closed geodesic contour. Using isothermal coordinates, we derive the Euler--Lagrange equation associated with the curvature functional and obtain Dirichlet and Neumann boundary conditions from the prescribed spatial boundary and its geodesic constraint. We then investigate a biharmonic approximation obtained by linearizing the nonlinear equation around a constant conformal factor.

The longer-term objective is to apply this analysis to the cells of a prescribed geodesic network, providing a framework for the computation of smooth surfaces from geometric data. Such a framework could contribute to CAD interfaces, sketch-to-3D modeling, and related problems in geometric reconstruction.

\subsection{Related work and our approach}
The reconstruction of surfaces from prescribed geodesic curves has been investigated in computer-aided geometric design. Sprynski et al.~\cite{GrenoblePeoplePaper} studied surface reconstruction via geodesic interpolation. Farouki, Szafran, and Biard examined existence conditions for Coons patches interpolating geodesic boundary curves~\cite{FaroukiCoons2009}, the construction of B\'ezier patches with B\'ezier curves as geodesic boundaries~\cite{FaroukiBezier2009}, and the construction and smoothing of triangular Coons patches with geodesic boundaries~\cite{FaroukiTriangular2010}. These works provide closely related interpolation techniques and geometric constraints.

A complementary line of work concerns variational surface energies. Brakke's \emph{Surface Evolver}~\cite{Brakke1992} provides a framework for numerical minimization of surface functionals; Osipov and Zelikin~\cite{OsipovZelikin2012} studied higher-order elastic energies, and Gruber, Toda, and Tran~\cite{GruberTodaTran2019} considered variations of curvature functionals, including generalized Willmore energies. Our approach combines a prescribed geodesic boundary constraint with minimization of total squared Gaussian curvature. We derive the associated equation and boundary data in isothermal coordinates as a step toward reconstructing patches of a geodesic network.

\section{Variational formulation and admissible surfaces}
Let $\Gamma\subset\mathbb R^3$ be a simple closed piecewise
$C^2$ spatial curve. We consider a nonempty class
$\mathcal A_\Gamma$ of compact, oriented, $C^2$ embedded
surfaces $S\subset\mathbb R^3$ with $\partial S=\Gamma$,
such that each smooth boundary arc is a geodesic of $S$.

We assume that each surface under consideration admits
a global isothermal chart $\phi:S^\circ\to\Omega$,
where $\Omega\subset\mathbb R^2$ is a bounded domain,
such that $\phi$ and its inverse extend as $C^2$ maps
with nondegenerate differentials to corresponding smooth
boundary arcs, mapping these arcs homeomorphically onto
one another. We further assume that the associated
conformal factor $f$ belongs to $C^4(\Omega)$ and
extends as a $C^2$ function to each smooth boundary arc.

The existence and regularity of minimizers of the
variational problem are not established here and are
assumed whenever needed.

The geometric objective is to minimize
\begin{equation}\label{eq:surface-energy}
\mathcal E(S)=\int_S K^2\,dA,
\end{equation}
where $K$ is Gaussian curvature. In isothermal coordinates on a planar domain $\Omega$, the induced metric is
\begin{equation}\label{eq:metric}
g=e^{2f(x,y)}(dx^2+dy^2).
\end{equation}
Since $K=-e^{-2f}\Delta f$ and $dA=e^{2f}\,dx\,dy$, the intrinsic functional becomes
\begin{equation}\label{eq:conformal-energy}
\mathcal E(f)=\int_\Omega e^{-2f}(\Delta f)^2\,dx\,dy.
\end{equation}

\subsection{Embeddability assumption}
The functional \eqref{eq:conformal-energy} is expressed intrinsically in terms of a conformal Riemannian metric. The intrinsic variational problem is well-defined for smooth conformal metrics independently of their realization as surfaces in $\mathbb R^3$. Realization of a computed metric by a sufficiently regular isometric embedding compatible with the prescribed spatial boundary and its geodesic constraints is an additional requirement, not established by the intrinsic Euler--Lagrange equation.

In the numerical experiments, we compute an approximate conformal factor and compare its associated Gaussian curvature with that of a known reference surface. We leave the question of existence of an isometric embedding of the resulting metric in $\mathbb R^3$, compatible with the prescribed spatial boundary and its geodesic constraints, to future study.

\subsection{Choice of isothermal coordinates}
Local existence of isothermal coordinates on a smooth surface is classical; see Chern~\cite{Chern1955} and Schlag~\cite{Schlag2014}. We assume, in addition, the existence of a suitable single global isothermal chart for the surface under consideration. This global assumption does not follow merely from local existence.

For numerical applications, orthogonal projection of the prescribed spatial boundary onto a suitably chosen affine plane may provide an initial approximation of the isothermal coordinate domain. This domain may subsequently be refined through an iterative search or optimization procedure to obtain a coordinate representation consistent with the geometric constraints. The theoretical justification and convergence analysis of such a procedure are left for future research.

\section{The Euler--Lagrange equation}
Let $\Omega\subset\mathbb R^2$ be a bounded domain and $f$ sufficiently regular, for example $f\in C^4(\Omega)\cap C^2(\overline\Omega)$.

\begin{proposition}[Euler--Lagrange equation]\label{prop:EL}
If $f$ is stationary for \eqref{eq:conformal-energy} under all smooth compactly supported variations in $\Omega$, then
\begin{equation}\label{eq:EL}
\Delta(e^{-2f}\Delta f)-e^{-2f}(\Delta f)^2=0\qquad\text{in }\Omega.
\end{equation}
Equivalently,
\begin{equation}\label{eq:ELexpanded}
\Delta^2f-4\nabla f\cdot\nabla(\Delta f)-3(\Delta f)^2+4|\nabla f|^2\Delta f=0.
\end{equation}
\end{proposition}
\begin{proof}
For $\varphi\in C_c^\infty(\Omega)$, set $f_\varepsilon=f+\varepsilon\varphi$. Differentiation yields
\[
\left.\frac{d}{d\varepsilon}\mathcal E(f_\varepsilon)\right|_{\varepsilon=0}
=2\int_\Omega e^{-2f}\bigl(\Delta f\,\Delta\varphi-(\Delta f)^2\varphi\bigr)\,dx\,dy.
\]
Integrating by parts twice, with no boundary terms because $\varphi$ is compactly supported, gives
\[
\delta\mathcal E(f)[\varphi]
=2\int_\Omega\bigl[\Delta(e^{-2f}\Delta f)-e^{-2f}(\Delta f)^2\bigr]\varphi\,dx\,dy.
\]
The fundamental lemma implies \eqref{eq:EL}. Expanding the Laplacian and multiplying by $e^{2f}$ yields \eqref{eq:ELexpanded}.
\end{proof}

\begin{remark}[Scope of the stationarity equation]
Proposition~\ref{prop:EL} concerns variations of the conformal factor on a fixed planar domain. Such variations need not arise from variations through isometrically embedded surfaces in $\mathbb R^3$ preserving the prescribed spatial boundary. Accordingly, the proposition establishes the Euler--Lagrange equation for the intrinsic fixed-domain problem; it does not assert that every minimizer of the original constrained embedded-surface problem satisfies this equation. The relation between these classes of variations is left for future study.
\end{remark}

\subsection{Linearization around a constant conformal factor}
Let $f=f_0+\varepsilon u$, where $f_0$ is constant. Substitution into \eqref{eq:ELexpanded} gives
\[
0=\varepsilon\Delta^2u-4\varepsilon^2\nabla u\cdot\nabla(\Delta u)
-3\varepsilon^2(\Delta u)^2+4\varepsilon^3|\nabla u|^2\Delta u.
\]
The first-order equation is
\begin{equation}\label{eq:biharmonic}
\Delta^2u=0.
\end{equation}
Thus the biharmonic equation is the linearization of the nonlinear Euler--Lagrange equation around a constant conformal factor, corresponding to a flat metric.

\section{Boundary conditions for geodesic contours}
Suppose $S\subset\mathbb R^3$ is a sufficiently smooth immersed surface with simple, closed, piecewise $C^2$ boundary $\Gamma$, each smooth boundary arc being a geodesic. Let $\phi:S\to\overline\Omega$ be a global isothermal chart whose inverse $X=\phi^{-1}$ extends as a $C^2$ regular parametrization up to each open smooth boundary arc, with conformal factor $f$ of class $C^1$ there. All boundary identities below are understood on the smooth arcs, away from corner points.

\begin{theorem}[Geodesic boundary conditions]\label{thm:BC}
Let $c:I\to\Gamma$ be a regular $C^2$ parametrization of a smooth boundary arc, and set $\gamma=\phi\circ c$, assumed regular and $C^2$. Let $n$ be the outward Euclidean unit normal to $\partial\Omega$, and let $\kappa_n$ be the signed Euclidean curvature of $\gamma$, defined using Euclidean arclength $s$ by $dT/ds=\kappa_n n$. Then, on each smooth boundary arc,
\begin{align}
 f(\gamma(t))&=\log\frac{\|c'(t)\|}{\|\gamma'(t)\|},\label{eq:dirichlet}\\
 \partial_n f&=\kappa_n.\label{eq:neumann}
\end{align}
\end{theorem}
\begin{proof}
Since $X(\gamma(t))=c(t)$ and the pullback metric is $e^{2f}(dx^2+dy^2)$,
\[
\|c'(t)\|^2=e^{2f(\gamma(t))}\|\gamma'(t)\|^2,
\]
which proves \eqref{eq:dirichlet}.

For \eqref{eq:neumann}, parameterize $\gamma$ by Euclidean arclength $s$ and put $T=\gamma'(s)$. The Levi--Civita connection of $g=e^{2f}\delta$ satisfies
\[
\nabla_T^gT=\gamma''(s)+2(\nabla f\cdot T)T-\nabla f.
\]
Since $\gamma$ is a geodesic, though $s$ need not be an affine parameter for $g$, its covariant acceleration is tangent to $\gamma$. Taking the Euclidean normal component gives
\[
0=\gamma''(s)\cdot n-\nabla f\cdot n=\kappa_n-\partial_n f.
\]
This proves \eqref{eq:neumann}.
\end{proof}

\begin{remark}[Corners]
The formulas hold on the smooth boundary arcs. At corners the outward normal and curvature need not be uniquely defined, and the Neumann condition is imposed arcwise. The Dirichlet traces must be compatible at corners when continuity of $f$ is required. A sufficiently regular conformal chart preserves the interior angles between adjacent arcs. Corner regularity for the fourth-order boundary-value problem requires separate analysis.
\end{remark}
\begin{remark}[Coordinate dependence]
The boundary data depend on both the prescribed spatial curves $c$ and their representations $\gamma$ in isothermal coordinates. A projected planar contour may be used as an initial numerical approximation, followed by refinement of the coordinate domain.
\end{remark}

\section{The conformal-factor boundary-value problem}
Fix a piecewise smooth domain $\Omega$, planar boundary parameterizations $\gamma_j:I_j\to\partial\Omega$, and corresponding spatial curves $c_j$. The interior stationarity equation and geodesic conditions give the system
\begin{equation}\label{eq:full-bvp}
\begin{cases}
\Delta(e^{-2f}\Delta f)-e^{-2f}(\Delta f)^2=0,&\text{in }\Omega,\\
f(\gamma_j(t))=\log\dfrac{\|c_j'(t)\|}{\|\gamma_j'(t)\|},&\text{on each smooth arc},\\
\partial_n f=\kappa_n,&\text{on each smooth arc}.
\end{cases}
\end{equation}
This system combines the intrinsic fixed-domain stationarity equation with boundary identities that hold for a realized surface having the prescribed geodesic arcs. It is not, without further assumptions on admissible embedded-surface variations, a necessary optimality system for the original extrinsic minimization problem. The formulation presupposes the chosen isothermal boundary representation.

\subsection{Linearized problem and its quadratic energy}
For numerical computation, the nonlinear equation is replaced by its linearization,
\begin{equation}\label{eq:linear-bvp}
\begin{cases}
\Delta^2 f=0,&\text{in }\Omega,\\
f=f_D,&\text{on }\partial\Omega,\\
\partial_n f=f_N,&\text{on the smooth boundary arcs},
\end{cases}
\end{equation}
where $f_D$ and $f_N$ are given by \eqref{eq:dirichlet}--\eqref{eq:neumann}. Once $f$ is computed, its metric curvature is $K_f=-e^{-2f}\Delta f$.

For fixed compatible boundary traces, \eqref{eq:linear-bvp} is the Euler--Lagrange equation of
\begin{equation}\label{eq:B}
\mathcal B(f)=\frac12\int_\Omega(\Delta f)^2\,dx\,dy.
\end{equation}
Indeed, variations $\varphi$ with zero Dirichlet and normal traces satisfy
\[
\delta\mathcal B(f)[\varphi]=\int_\Omega\Delta f\,\Delta\varphi
=\int_\Omega\Delta^2f\,\varphi.
\]
More precisely, suppose $\Omega$ is a bounded domain for which the standard $H^2_0(\Omega)$ coercivity estimate holds, and the prescribed traces admit an $H^2(\Omega)$ lifting $F$. The admissible class is then $F+H^2_0(\Omega)$. The quadratic functional is coercive on this affine class and has a unique minimizer by the direct method: if two minimizers existed, their difference would have zero Laplacian and zero Dirichlet trace, hence would vanish. The Euler--Lagrange equation holds weakly, and classically when additional regularity is available. For piecewise smooth domains, the required lifting and corner compatibility are separate assumptions.

\begin{proposition}[Energy approximation]\label{prop:energy}
Let $\mathcal A$ be a nonempty common admissible class of sufficiently regular conformal factors on a fixed domain $\Omega$, with identical compatible Dirichlet and Neumann data. Suppose $f_B,f_E\in\mathcal A$ minimize $\mathcal B$ and $\mathcal E$, respectively, over this same class. If, for some constant $f_0$ and $\delta\geq0$,
\[
\|f_B-f_0\|_{L^\infty(\Omega)}\leq\delta,
\qquad
\|f_E-f_0\|_{L^\infty(\Omega)}\leq\delta,
\]
then
\begin{equation}\label{eq:ratio}
\mathcal E(f_B)\leq e^{4\delta}\mathcal E(f_E).
\end{equation}
\end{proposition}
\begin{proof}
For either $f=f_B$ or $f=f_E$, the assumed bound gives
\[
2e^{-2f_0-2\delta}\mathcal B(f)\leq\mathcal E(f)
\leq2e^{-2f_0+2\delta}\mathcal B(f).
\]
Consequently,
\[
\mathcal E(f_B)\leq2e^{-2f_0+2\delta}\mathcal B(f_B)
\leq2e^{-2f_0+2\delta}\mathcal B(f_E)
\leq e^{4\delta}\mathcal E(f_E).
\]
\end{proof}
\begin{remark}
The estimate concerns conformal factors on the same fixed coordinate domain with the same boundary data. The bounds on $f_B$ and $f_E$ are hypotheses, not consequences of the geodesic boundary conditions. The estimate does not by itself compare all embedded surfaces spanning the spatial contour.
\end{remark}

\subsection{Relation to surface reconstruction}
The coordinate domain and planar representation of the boundary must be selected or approximated before solving either boundary-value problem. Whether the resulting metric admits a sufficiently regular isometric embedding into $\mathbb R^3$ compatible with the prescribed spatial boundary and its geodesic constraints is not addressed here. Constructing and validating such an embedding is a separate stage of the proposed numerical program.

\section{Numerical experiments on an ellipsoid}
\subsection{Geodesic network and selection of cells}
\begin{figure}[htbp]
\centering
\includegraphics[width=.94\linewidth]{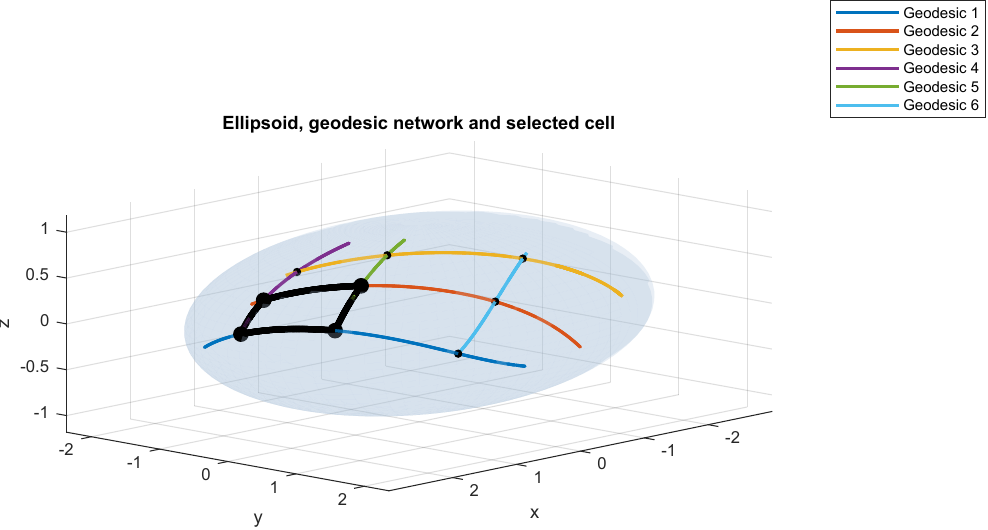}
\caption{Reference ellipsoid and the numerically integrated network of six geodesics. Intersections and a selected quadrilateral cell are highlighted.}
\label{fig:network}
\end{figure}

We consider the reference ellipsoid
\begin{equation}
\frac{X^2}{9}+\frac{Y^2}{4}+Z^2=1,
\qquad
\mathbf X(u,v)=(3\cos v\cos u,\,2\cos v\sin u,\,\sin v).
\end{equation}
Its geodesics are obtained by numerical integration of the geodesic equations for the induced metric. The resulting network contains six geodesics and four selected quadrilateral cells, with boundaries formed by geodesic segments (Figure~\ref{fig:network}). The cells are bounded, respectively, by geodesics $(1,2,4,5)$, $(1,2,5,6)$, $(2,3,4,5)$, and $(2,3,5,6)$.

The ellipsoid is chosen because its geometry is explicitly known while its Gaussian curvature varies over the surface. This permits comparison with independently computed reference curvature. We do \emph{not} assume that the ellipsoidal patch minimizes the squared-Gaussian-curvature energy among admissible surfaces spanning the same geodesic contour. Thus, a discrepancy from the ellipsoid is not necessarily an error relative to the variational minimizer.

\subsection{Approximate isothermal charts and boundary data}
\begin{figure}[htbp]
\centering
\includegraphics[width=.94\linewidth]{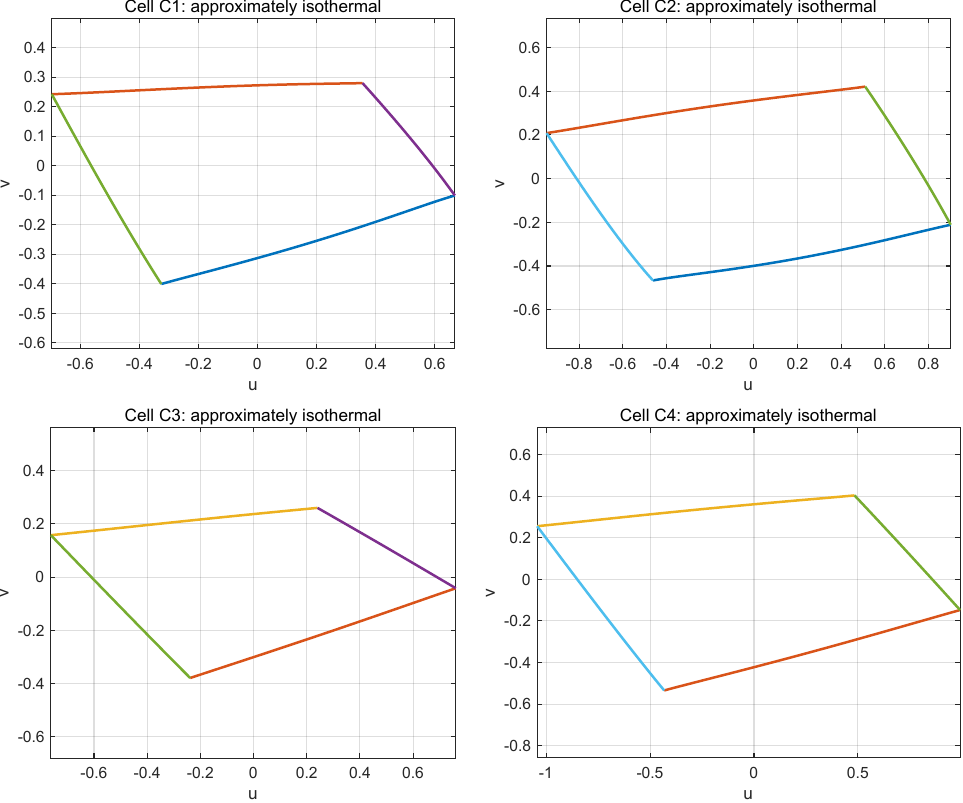}
\caption{The four computational domains in their numerically constructed approximately isothermal coordinates.}
\label{fig:domains}
\end{figure}

For each cell, we fit an affine plane to its four vertices and express the reference patch locally as a graph over the fitted plane. The resulting metric is used to construct an approximately isothermal chart by solving a Laplace--Beltrami problem for one coordinate and recovering an approximate harmonic conjugate. The relationship between harmonic functions, their local harmonic conjugates, and conformal coordinates is classical (see Schlag~\cite{Schlag2014}); computational conformal geometry provides broader context for such constructions (see Gu, Luo, and Yau~\cite{GuLuoYau2009}). In our implementation, the conjugate coordinate is recovered approximately by least squares, rather than by an exact analytic conjugation. The fitted plane is only an intermediate representation; the conformal-factor problem is formulated in the approximate isothermal coordinates. Figure~\ref{fig:domains} shows the resulting four planar domains.

Writing $w(s)$ for a boundary geodesic in the approximate isothermal coordinates, parametrized by physical arc length, the prescribed boundary data are calculated from
\begin{equation}
f_D(w(s))=-\log|w'(s)|,\qquad \partial_n f=\kappa_{\mathrm{out}},
\end{equation}
where $\kappa_{\mathrm{out}}$ is the signed Euclidean curvature of the planar boundary with respect to the outward normal. Because the computed coordinates are only approximately isothermal, the boundary values obtained from these formulas approximate the data of the exact isothermal-coordinate theorem; they are not exact boundary data for the reference metric. The reported mean conformal distortion values for the four charts range from approximately $1.00006$ to $1.00033$. Small measured conformal distortion does not, by itself, provide an error bound for the recovered curvature. These tests use the known ellipsoid metric to construct the chart; they do not establish a chart-recovery method from spatial boundary data alone.

\subsection{Biharmonic computation and curvature comparison}
We solve the biharmonic approximation $\Delta^2 f=0$ with the computed Dirichlet and Neumann boundary data using the implementation of a fourth-order finite-difference scheme developed by A.~Ditkowski. Ditkowski's doctoral dissertation~\cite{Ditkowski1997} provides foundational background on bounded-error finite-difference schemes for initial-boundary-value problems on complex domains. A closely related embedded finite-difference approach, including a decomposition of the biharmonic operator and treatment of boundary conditions on complex domains, is developed in Chapter~4 of Livneh's thesis, supervised by Ditkowski~\cite{Livneh2015}. For a classical alternative formulation, see Ciarlet and Raviart~\cite{CiarletRaviart1974}. The reconstructed curvature is calculated as $K_h=-e^{-2f_h}\Delta f_h$ in the approximate isothermal chart.

\begin{figure}[p]
\centering
\includegraphics[width=\textwidth,height=.83\textheight,keepaspectratio]{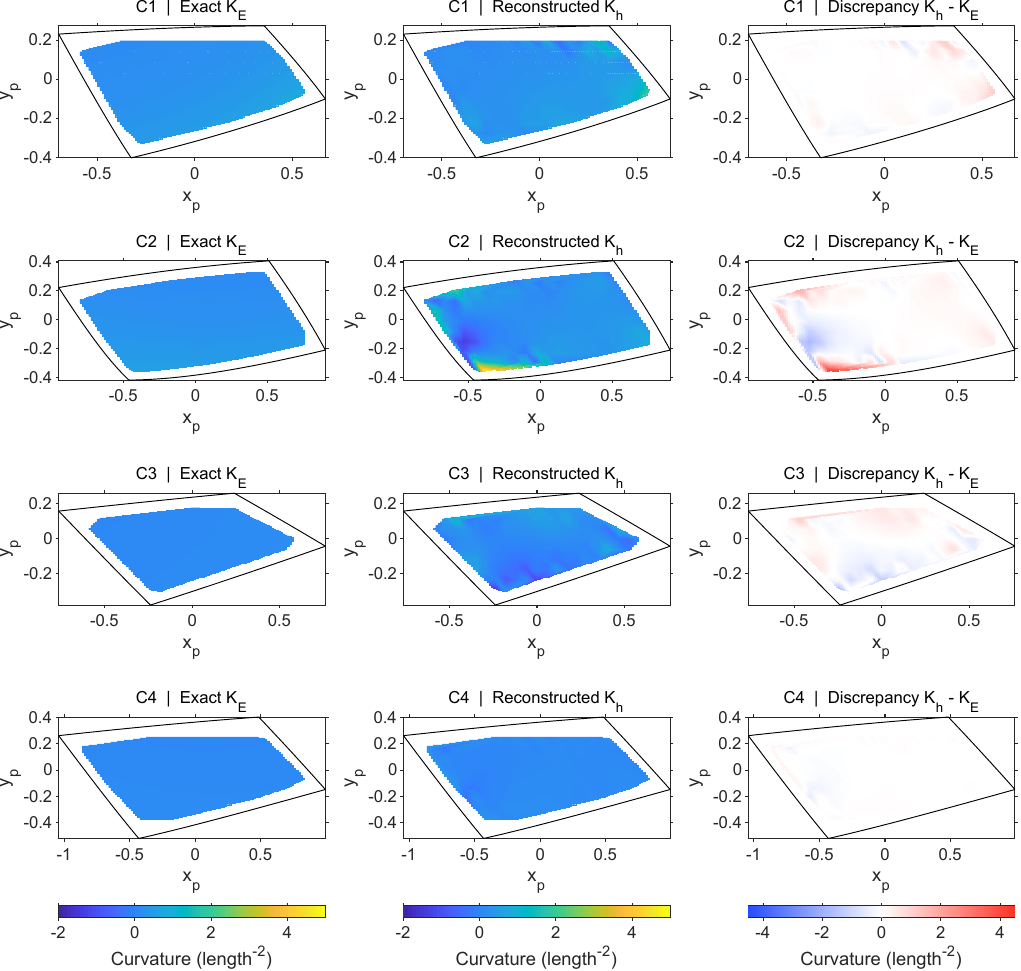}
\caption{Curvature comparisons for the four geodesic cells at $N=31$: reference ellipsoid curvature, reconstructed curvature, and pointwise discrepancy $K_h-K_E$ at corresponding comparison points. The displayed curvature differences are not errors relative to a known variational minimizer.}
\label{fig:curvature}
\end{figure}

Figure~\ref{fig:curvature} compares the reference curvature, reconstructed curvature, and their pointwise discrepancy for all four cells at the saved grid configuration $N=31$. The reference and reconstructed values are compared at corresponding physical points. The results exhibit partial consistency with the ellipsoidal reference geometry, with substantial differences among the cells and localized discrepancies near their boundaries.

For comparison points $p_1,\ldots,p_M$, we define the root-mean-square (RMS) curvature discrepancy by
\begin{equation}
\operatorname{RMS}(K_h,K_E)=\left(\frac1M\sum_{j=1}^M\bigl(K_h(p_j)-K_E(p_j)\bigr)^2\right)^{1/2}.
\end{equation}
Table~\ref{tab:curvature} reports the RMS and maximum absolute discrepancies using two sampling conventions. The common-point RMS is evaluated at a fixed set of physical comparison points on each ellipsoidal patch, allowing comparison across computational configurations. The native-grid RMS is evaluated directly at the interior nodes of the numerical grid in the approximate isothermal coordinates, mapped to their corresponding physical points. These samplings differ, particularly near the boundary, and can therefore yield different discrepancy statistics.
\begin{table}[htbp]
\centering
\caption{Curvature discrepancies for the four cells at $N=31$, from the saved numerical results.}
\label{tab:curvature}
\begin{tabular}{ccccc}
\toprule
Cell & Common RMS & Common maximum & Native RMS & Native maximum\\
\midrule
C1 & 0.1520 & 1.1840 & 0.2272 & 1.2008\\
C2 & 0.4737 & 4.1339 & 0.5858 & 4.4315\\
C3 & 0.3022 & 1.5458 & 0.3890 & 1.8587\\
C4 & 0.0670 & 0.4740 & 0.0931 & 0.5207\\
\bottomrule
\end{tabular}
\end{table}

The largest discrepancies on the native grids occur approximately $1.04$--$1.46$ grid spacings from the boundary. This suggests sensitivity to the boundary treatment, although the numerical evidence does not establish a unique cause. The results are presented as geometric comparisons, not as a grid-convergence study. In particular, the observed discrepancy can reflect the approximation of isothermal coordinates, the replacement of the nonlinear Euler--Lagrange equation by its biharmonic linearization, and numerical discretization of the fourth-order boundary-value problem. These contributions have not been isolated. Even convergence of the numerical biharmonic solver would not imply convergence of its curvature to the ellipsoid curvature: the reference ellipsoid is not known to be the variational minimizer for these contours.

\section{Discussion and future directions}
The analysis provides a variational formulation for surfaces spanning prescribed geodesic contours, together with the conformal-factor equation and its boundary conditions. The ellipsoid experiments illustrate the construction of approximately isothermal charts on four cells and permit comparisons with independently known curvature. These comparisons are informative about geometric consistency but do not establish recovery of the variational minimizer or numerical convergence.

A central next step is to construct suitable coordinate domains and boundary data when the underlying surface is unknown, rather than using the reference ellipsoid metric. Further questions include the existence and regularity of compatible surface embeddings, the consistency of adjacent reconstructed cells, and more systematic numerical validation of the fourth-order boundary-value solver. Experiments on geometries with greater curvature variation may also help assess the scope of the method.

\section*{Acknowledgments}
I wish to thank Prof.~Michel Bercovier of the Hebrew University for insightful discussions concerning this project.

I am grateful to Professors Franco Brezzi, Peter Monk, and Michael Neilan for their valuable suggestions and assistance in exploring numerical methods during earlier stages of this research.

I am deeply grateful to Prof.~Brittany Froese Hamfeldt for her interest in this project and for generously sharing her MATLAB code with me. Although her solver was not used in the numerical experiments reported here, it was valuable in the earlier development of this work.

I am particularly grateful to Prof.~Adi Ditkowski for guiding me through the implementation of his numerical scheme, which is used in the numerical experiments reported here.

I acknowledge the use of ChatGPT and Codex in the preparation of this manuscript.

\end{document}